# On the nature of monetary and price inflation and hyperinflation

Laurence Francis Lacey

Lacey Solutions Ltd, Skerries, County Dublin, Ireland

Fri 13 August 2021

# On the nature of monetary and price inflation and hyperinflation

Laurence Francis Lacey

Lacey Solutions Ltd, Skerries, County Dublin, Ireland

Fri 13 August 2021

## Abstract

Monetary inflation is a sustained increase in the money supply than can result in price inflation, which is a rise in the general level of prices of goods and services. The objectives of this paper were to develop economic models to (1) predict the annual rate of growth in the US consumer price index (CPI), based on the annual growth in the US broad money supply (BMS), the annual growth in US real GDP, and the annual growth in US savings, over the time period 2001 to 2019; (2) investigate the means by which monetary and price inflation can develop into monetary and price hyperinflation.

The hypothesis that the annual rate of growth in the US CPI is a function of the annual growth in the US BMS minus the annual growth in US real GDP minus the annual growth in US savings, over the time period investigated, has been shown to be the case. However, an exact relationship required the use of a non-zero residual term. A mathematical statistical formulation of a hyperinflationary process has been provided and used to quantify the period of hyperinflation in the Weimar Republic, from July 1922 until the end of November 1923.

# 1. Introduction

Monetary inflation is a sustained increase in the money supply of a country (or currency area) and it is likely to result in price inflation, which is usually just called "inflation", which is a rise in the general level of prices of goods and services [1]. The consumer price index (CPI) is a common measure of price inflation [2]. While there is general agreement among economists that there is a causal relationship between monetary inflation and price inflation, there is no general agreement on the exact relationship between the two [1]. Hyperinflation is rapidly rising price inflation, typically measuring more than 50% per month [3]. Broad money includes both notes and coins, but also other forms of money, which can easily be converted into cash. It is the most inclusive method of calculating a given country's money supply [4].

The objectives of this paper are to investigate:

(1) the hypothesis that the annual rate of growth in the US consumer price index (CPI) is a function of the annual growth in the US broad money supply (BMS) minus the annual growth in US real GDP minus the annual growth in US savings, over the time period 2001 to 2019, with 2001 as the reference year (time = 0).

(2) the means by which monetary and price inflation can develop into monetary and price hyperinflation.

## 2. Methods

### 2.1 Statistical Methodology

An information entropy statistical methodology has been developed for investigating expansionary processes, in with full details of the methodology can be found [5,6,7]. For an exponential expansionary process, the methodology provides a rate-constant (λ) for the exponential growth of the process (G(t)) and the associated information entropy for the time series under investigation [5,7]. This can be expressed as follows:

$$G(t) = \exp(\lambda \, x \, t)$$

and,

$$\text{Info Ent } (G(t)) = \lambda \, x \, t$$

While information entropy has no units, at any given time, it is related to the average growth rate ($r$), where:

$$r = \exp(\lambda) - 1 \quad \text{and} \quad \lambda = \log_e(1 + r)$$

Consequently, the information entropy of an expansionary process can be considered to be related to the "velocity" of the growth of a time series of values [5,7]. Henceforth, in the interests of nomenclature brevity, the following nomenclature will be adopted:

$$vG(t) = \text{Info Ent } (G(t))$$

where, v relates to the "velocity" of growth of any given time series, G(t).

The hypothesis that this paper will investigate, for the US time series over the period 2001 to 2019, is:

$$vCPI(t) = vBMS(t) - vGDP(t) - vSAV(t)$$

where CPI is the consumer price index, BMS is the broad money supply, and SAV are average savings.

## 2.2 US Time Series Data

The sources of the annual US time series data, over the period 2001 to 2019, are: CPI [8], BMS [9] real GDP [10]. The annual savings data are those estimated for American households. These savings data were reported for the years 2001, 2004, 2007, 2010, 2013, 2016, 2019 only [11]. Where required, the annual savings estimates for the years not available were imputed using the regression fit obtained using the available years of data.

The time series data, over the time period 2001 to 2019, with 2001 as the reference year (time = 0), were expressed relative to the value given in year 2001. The rate-constant (λ) estimates for each time series, together with the 95% confidence intervals, and coefficients of determination ($R^2$) of the regression fits were obtained from a linear regression (intercept = 0) of the natural log-transformed time series data versus time, with time = 0 for the reference year 2001, using Python (version 3.9.2) statsmodels package [12].

## 2.3 Hyperinflation Model

A mathematical statistical characterisation of a hyperinflation process is given in the Supplementary Appendix, building on the original characterisation of a process with a mono-exponentially increasing sample space [5]. As is shown in the Supplementary Appendix, an exponential growth process becomes a double exponential growth process, when a second exponential growth process occurs, at some point in time ($t^*$), on top of the initial exponential growth process.

When $t < t^*$,

$$vG(t) = \lambda_1 \: x \: t$$

When $t \geq t^*$,

$$vG(t - t^*) = (\lambda_1 \: x \: t^*)^{exp(\lambda_2 \: x \: (t-t^*))}$$

$$log_e(vG(t - t^*)) = (log_e(\lambda_1 \: x \: t^*)) + (\lambda_2 \: x \: (t - t^*))$$

While the information entropy of an expansionary process can be considered to be related to the "velocity" of the expansion, the natural log of the information entropy of an expansionary process can be considered to be related to the "acceleration" of the expansion.

The double exponential nature of some cases of monetary hyperinflation has been investigated [13] and there has been an exploratory investigation of a double exponential hyperinflationary process, using the information entropy methodology [6].

All plots of the data analysis given below were obtained using Microsoft Excel 2019, 32-bit version.

## 3. Results

### 3.1 Characterisation of Price Inflation

A plot of the natural log-transformed US BMS time series over the period 2001 to 2019 is given in Figure 1. The year 2001 is the reference year (time = 0), and the BMS data were expressed relative to the value obtained in year 2001. The corresponding plots for the natural log-transformed US CPI time series is given in Figure 2; the natural log-transformed US real GDP time series is given in Figure 3; the natural log-transformed average American household savings time series is given in Figure 4

Figure 1: Plot of the natural log-transformed US BMS time series over the period 2001 to 2019, with 2001 as the reference year (time = 0)

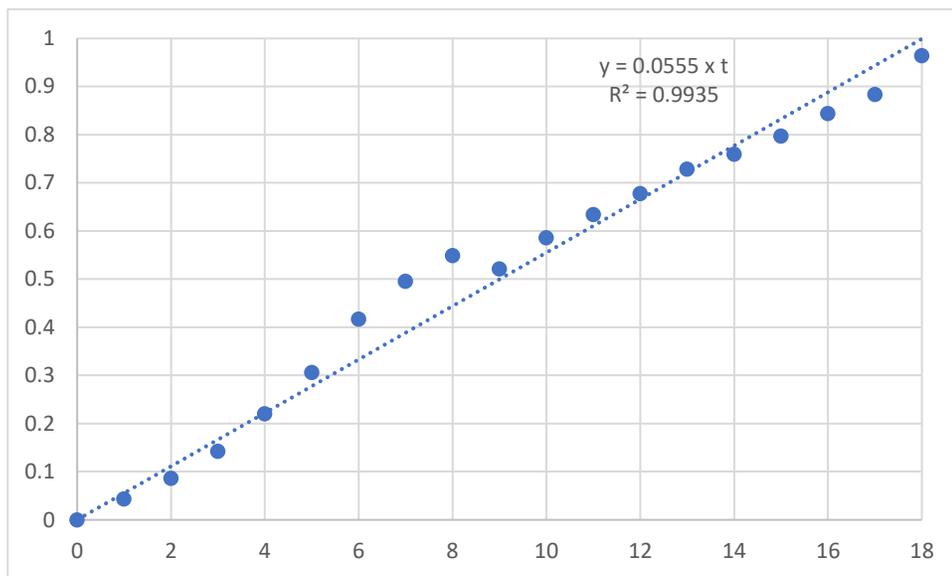

The average annual rate of growth of the US BMS, over the period 2001 to 2019, was 5.7%, with:

$$vBMS(t) = 0.0555 \text{ x t}$$

Figure 2: Plot of the natural log-transformed US CPI time series over the period 2001 to 2019, with 2001 as the reference year (time = 0)

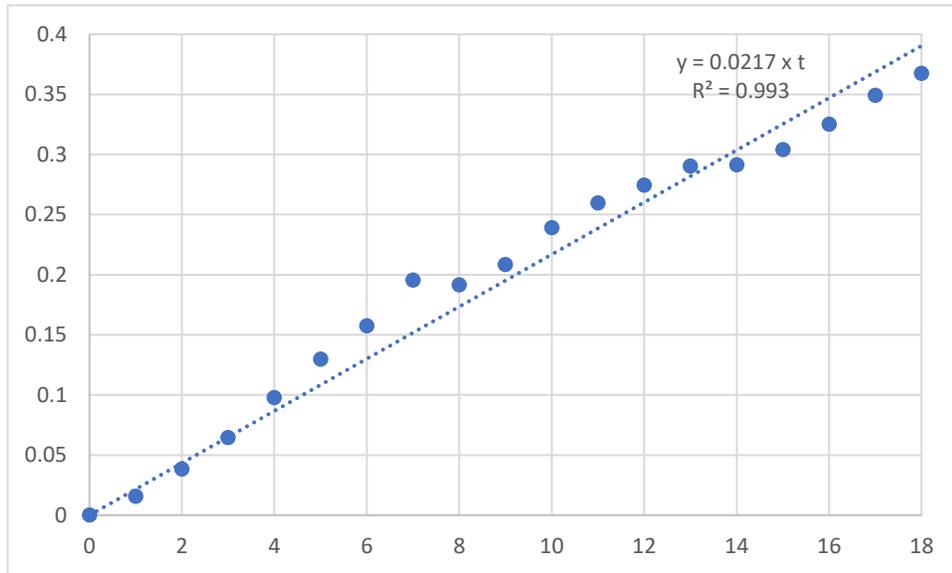

The average annual rate of US price inflation as measured by CPI, over the period 2001 to 2019, was 2.2%, with:

$$vCPI(t) = 0.0217 \text{ x t}$$

Figure 3: Plot of the natural log-transformed US real GDP time series over the period 2001 to 2019, with 2001 as the reference year (time = 0)

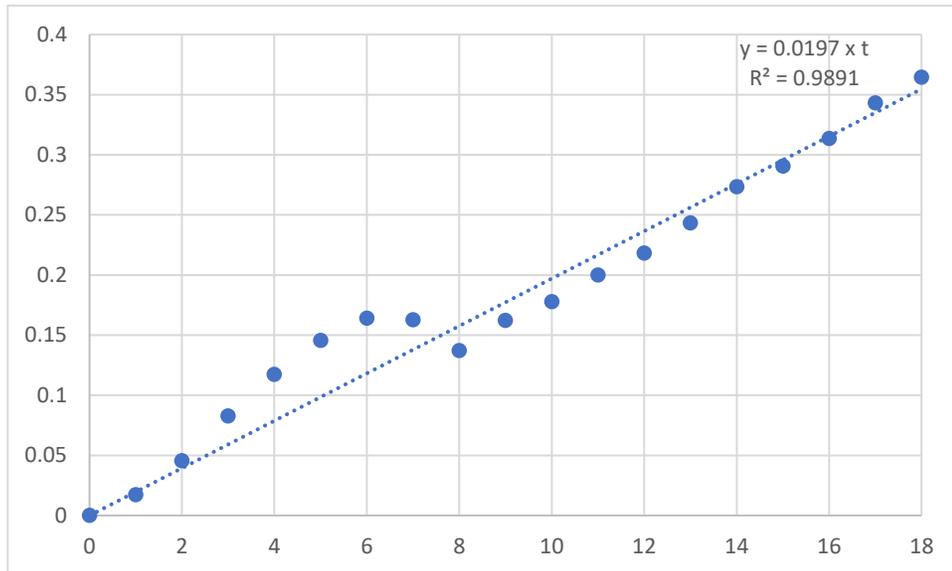

The average annual rate of growth of the US real GDP, over the period 2001 to 2019, was 2.0%, with:

$$vGDP(t) = 0.0197 \text{ x } t$$

Figure 4: Plot of the natural log-transformed average annual American household savings over the period 2001 to 2019, with 2001 as the reference year (time = 0)

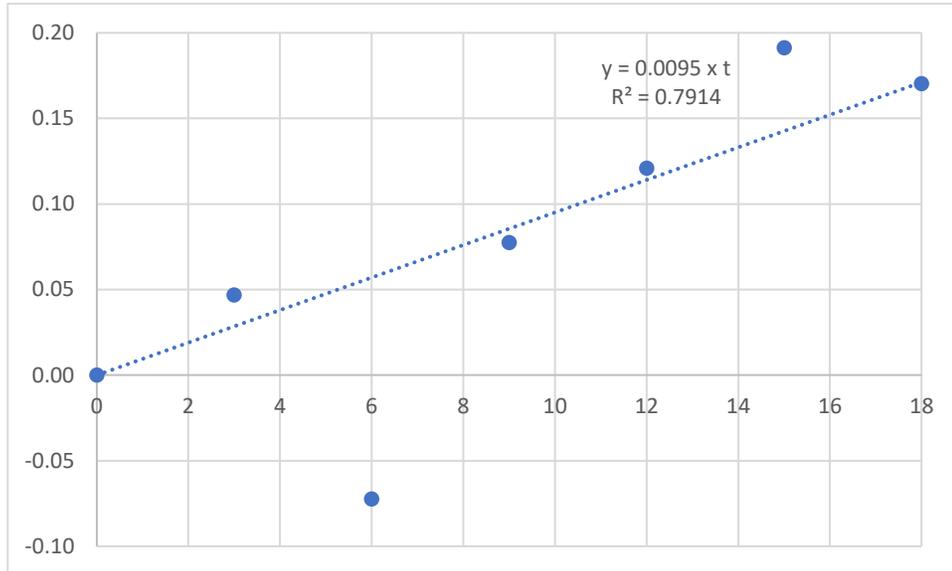

The average annual rate of growth of American household savings, over the period 2001 to 2019, was 1.0%, with:

$$vSAV(t) = 0.0095 \text{ x t}$$

The linear regression fits are summarised in Table 1.

Table 1: Rate-constant (λ) estimates for US CPI, US BMS, real US GDP, and US household savings (SAV), over the period 2001 to 2019, together with the corresponding 95% confidence intervals and $R^2$ values

| Economic Index | Estimated rate-constant (λ) | 95% confidence interval | $R^2$ | df Residuals |
|---|---|---|---|---|
| BMS | 0.0555 | 0.053 to 0.058 | 99.4% | 18 |
| CPI | 0.0217 | 0.021 to 0.023 | 99.3% | 18 |
| GDP | 0.0197 | 0.019 to 0.021 | 98.9% | 18 |
| SAV | 0.0095 | 0.005 to 0.014 | 79.1% | 6 |

BMS, broad money supply; CPI, consumer price index; GDP, (real) gross domestic product; SAV, average American household savings; df, degrees of freedom

Henceforth, the annual US household savings estimates for the years not available were imputed, using the regression fit obtained from the available years of data.

The residual time series (RES(t)) was obtained from the relationship:

$$RES(t) = vBMS(t) - vGDP(t) - vSAV(t) - vCPI$$

This residual time series is plotted in Figure 5.

Figure 5: Plot of the residual time series (RES(t)) over the period 2001 to 2019, with 2001 as the reference year (time = 0)

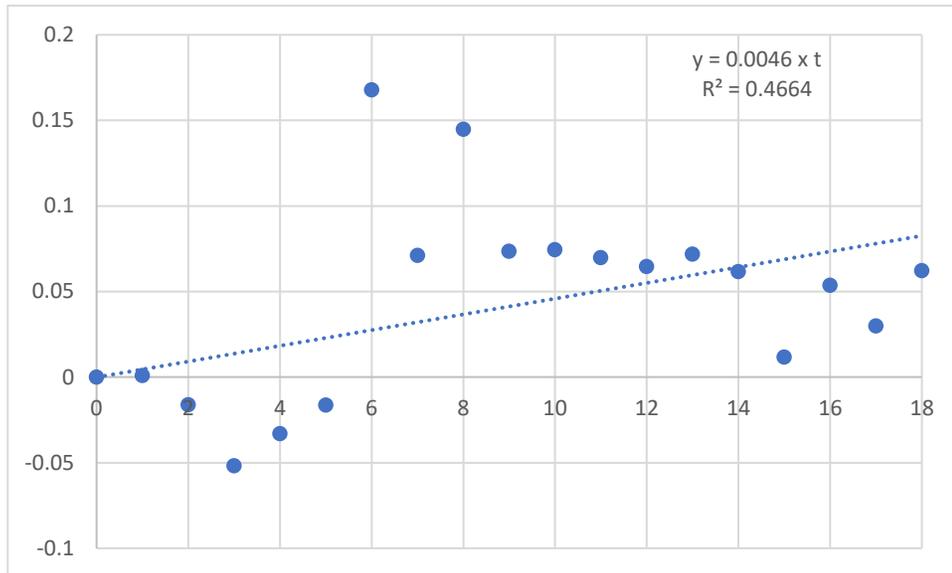

The average annual rate of growth of the residual time series (RES(t)), over the period 2001 to 2019, was 0.5%, with:

$$RES(t) = 0.0046 \text{ x t}$$

This gives:

$$vCPI(t) = vBMS(t) - vGDP(t) - vSAV(t) - RES(t)$$

Further detail of the residual time series (RES(t)), over the period 2001 to 2019, are given in Table 2.

Table 2: Rate-constant (λ) estimate for the residual time series (RES(t)) over the period 2001 to 2019, together with the corresponding 95% confidence interval and $R^2$ value.

| Economic Index | Estimated rate-constant (λ) | 95% confidence interval | $R^2$ | df Residuals |
|---|---|---|---|---|
| RES | 0.0046 | 0.002 to 0.007 | 46.6% | 18 |

RES, residual time series; df, degrees of freedom

## 3.2 Characterisation of Hyperinflation in the Weimar Republic (1922 to 1923)

The monetary and price hyperinflation of in the Weimar Republic between the middle of 1920 to the end of November 2023 is investigated in this paper and the data analysed are based on the value of one gold Mark in paper Marks over this time period (obtained from [14]) and are plotted in Figure 6 (semi-log plot).

Figure 6: Semi-log plot of the value of one gold Mark in paper Marks over the period from the middle of 1920 to the end of November, 1923 (in months), with time t=0 being the middle of 1920 (obtained from [14]).

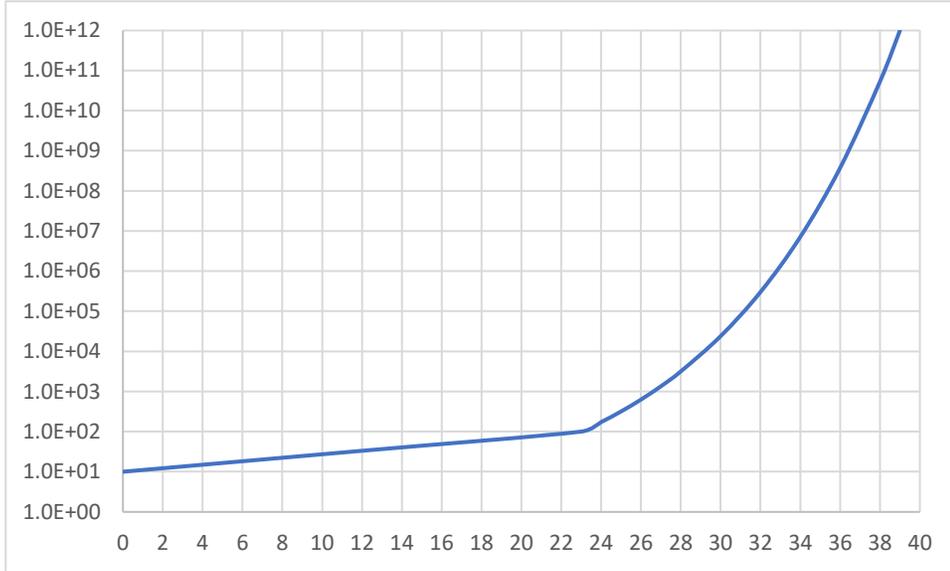

This is a clear case of hyperinflation that is double exponential in nature [6,13]. In the nomenclature of this paper, $t^*$ occurred when t = 23 months (July 1922).

When t < 23,

$$\text{vPI}(t) = (0.1001 \, x \, t) + \log_e(10)$$

where, PI(t) is price inflation, with a growth rate of 10.4% per month (inflation velocity), and where 10 is the value of PI(t) at t = 0.

When t ≥ 23,

$$\text{vPI}(t-23) = ((0.1001 \, x \, 23) + \log_e(10))^{exp(0.112 \, x \, (t-23))}$$

$$\log_e(\text{vPI}(t-23)) = \log_e((0.1001 \, x \, 23) + \log_e(10)) + (0.112 \, x \, (t-23)), \text{giving}$$

$$log_e(vPI(t-23)) = (0.112 \, x \, (t-23)) + 1.527$$

During this period of hyperinflation, the <u>acceleration</u> in price inflation was 11.8% per month.

## 4.  Discussion

The consumer price inflation model described is intended to investigate the consumer price index (CPI), over a period of years and not over a period of months, in order to investigate the longer-term relationships involved. In contrast, the hyperinflation model described is intended to investigate the increase in nominal prices, over a period of months, because of the accelerating nature of the decline in the purchasing power of the fiat currency being studied. Unlike in the case of hyperinflation, under "ordinary" circumstances, the growth in price inflation has a velocity, but the acceleration can be ignored because it is approximately constant, i.e., price inflation increases at an approximate constant rate with time.

The hypothesis that the annual rate of growth in the US CPI is a function of the annual growth in the US broad money supply minus the annual growth in US real GDP minus the annual growth in US savings, over the time period 2001 to 2019, has been shown to be the case. However, an exact relationship required the use of a non-zero residual term, RES(t):

$$vCPI(t) = vBMS(t) - vGDP(t) - vSAV(t) - RES(t)$$

This may be the case for a couple of reasons:

(1) The measure of savings used was the estimated average annual American household savings [11]. While, this might be expected to be technically challenging to calculate accurately, it only pertains to households. It might be expected that in addition to average annual American household savings, there should also exist annual American <u>non-household</u> savings. Therefore, a limitation of this paper is that American household savings were used as a measure of overall savings in the US economy.

(2) There may be an additional factor that needed to be included in the model, other than savings.

Savings were included in the model because savings can be used, if required / desired to purchase goods and/or services, or not, as the case might be, and as a result might be expected to influence the CPI. An alternative approach might be to use another broad measure of the money supply that excludes savings.

The model includes the terms, BMS and GDP. So also does the concept of the "velocity of money" [15], in which:

$$"velocity\ of\ money" = GDP/Money\ Supply$$

This equation can be rearranged as follow:

$$log_e(BMS) - log_e(GDP) = -\ log_e("velocity\ of\ money")$$

However, for the concept of the "velocity of money" to be relevant to this paper, it would need to be expressed in terms of its constant rate of change with time, as follows:

$$vBMS(t) - vGDP(t) = -\log_e(\text{velocity of money}(t))$$

Should the CPI model be valid (subject to the caveat discussed above in relation to savings) such that:

$$vCPI(t) = vBMS(t) - vGDP(t) - vSAV(t)$$

this would predict a few different economic scenarios:

(1) When there is no (or very little) growth in GDP or savings ("stagflation scenario"):

$$vCPI(t) = vBMS(t)$$

(2) When the growth in savings is sufficiently high, the growth in CPI could turn negative ("deflation scenario").

(3) When there is negative growth in GDP ("recession scenario"), the growth in CPI could be greater than the growth in BMS, should there be no "rebalancing" change in the growth of BMS.

A mathematical statistical formulation of a hyperinflationary process has been provided and used to quantify the period of hyperinflation in the Weimar Republic from July 1922 until the end of November 1923. Why did this hyperinflationary happen and why was it allowed to occur by the Government and central bankers of the Weimar Republic?

As outlined in [14], following World War 1 and the Treaty of Versailles (which came into effect on January 10, 1920), the strategy that Germany used to pay war reparations was the mass printing of bank notes to buy foreign currency, which was then used to pay reparations, but which also caused inflation of the paper mark. After Germany failed to pay France an instalment of reparations on time in late 1922, French and Belgian troops occupied the Ruhr valley, Germany's main industrial region, in January 1923. The German government's response was to order a policy of passive resistance in the Ruhr, akin to a general strike, to protest the occupation, while providing financial assistance to the striking workers by printing more and more banknotes, with the result that the paper mark became increasing worthless throughout 1923.

The mathematical statistical formulation of a hyperinflationary process could be extended beyond two sequential exponential expansions to include any arbitrary number (n) of sequential exponential expansions. However, while this may have theoretical interest, it is not a phenomenon that has yet occurred in real-world economics, to the knowledge of the author.

## 5. Conclusion

The hypothesis that the annual rate of growth in the US consumer price index is a function of the annual growth in the US broad money supply minus the annual growth in US real GDP minus the annual growth in US savings, over the time period 2001 to 2019, has been shown to be the case. However, an exact relationship required the use of a non-zero residual term. A mathematical statistical formulation of a hyperinflationary process has been provided and used to quantify the period of hyperinflation in the Weimar Republic from July 1922 until the end of November 1923. The main mathematical difference between inflation and hyperinflation is that the former is a process with a constant rate of increase, at a constant acceleration, whereas the latter is a process with a constant rate of acceleration.

## Supplementary materials

There is an Appendix providing the mathematical statistical characterization of a hyperinflationary process.

## Acknowledgements

No financial support was received for any aspect of this research.

[14] "Hyperinflation in the Weimar Republic". Accessed on 18 May 2021 from Wikipedia.com. url: https://en.wikipedia.org/wiki/Hyperinflation_in_the_Weimar_Republic

[15] "Velocity of Money". Accessed on 13 April 2021 from Investopedia. url: https://www.investopedia.com/terms/v/velocity.asp

# Appendix: Characterization of a hyperinflationary process – one with an initial exponential expansion of the sample space followed a subsequent second exponential expansion of the sample space

A process with an exponentially increasing sample space (s) has been characterised [5]. When the exponential expansion occurs with time (t), it has been shown [5] that:

$$s(t) = \exp(\lambda \times t)$$

$$H(t) = \lambda \times t$$

where, s(t) is the size of the sample space with time, H(t) is the information entropy of the expansionary process with time, and λ is exponential rate constant.

Let the initial exponential expansion occur until time, $t^*$, with rate constant = $\lambda_1$.

After this time (t ≥ $t^*$), the exponential expansion of the sample space occurs with rate constant = $\lambda_2$. Let $s_1(t)$ be the initial expansion of the sample space and $s_2(t)$ be the subsequent expansion for time (t ≥ $t^*$).

Thus, for, t < $t^*$:

$$s_1(t) = \exp(\lambda_1 \times t) \tag{1}$$

$$H_1(t) = \lambda_1 \times t \tag{2}$$

Setting:

$$s_2(t^*) = \log_e(s_1(t^*)) \times \frac{1}{\lambda_1 \times t^*} = 1 \tag{3}$$

For time (t ≥ t*):

$$s_2(t - t^*) = \exp(\lambda_2 \; x \; (t-t^*))$$

$$H_2(t - t^*) = \lambda_2 \; x \; (t - t^*)$$

Substitution equation (3) for $s_2(t^*)$, gives the overall expansion in terms of $s_1(t - t^*)$, for t ≥ t*:

$$s_1(t - t^*) = \exp\bigl(\lambda_1 \; x \; t^* \; x \; \exp(\lambda_2 \; x \; (t - t^*))\bigr) = \exp(\lambda_1 \; x \; t^*)^{\exp(\lambda_2 \; x \; (t-t^*))} \quad (4)$$

$$H_1(t - t^*) = (\lambda_1 \; x \; t^*)^{\exp(\lambda_2 \; x \; (t-t^*))} \quad (5)$$

$$\log_e\bigl(H_1(t - t^*)\bigr) = (\log_e(\lambda_1 \; x \; t^*)) + (\lambda_2 \; x \; (t - t^*)) \quad (6)$$